\documentclass[reprint,amsmath,amssymb,aps]{revtex4-2}

\usepackage{amsmath,amssymb,amsfonts}
\usepackage{amsthm}
\usepackage[switch]{lineno}
\usepackage{mathrsfs}
\usepackage{subcaption}
\usepackage{tabularx}
\usepackage{graphicx}
\usepackage{xcolor}

\begin{document}

\title{Arbitrage equilibrium and the emergence of universal microstructure\\ in deep neural networks}

\author{Venkat Venkatasubramanian}%
\email{venkat@columbia.edu}
\affiliation{Complex Resilient Intelligent Systems Laboratory, Department of Chemical Engineering, Columbia University, New York, NY 10027}

\author{N. Sanjeevrajan}
\affiliation{Complex Resilient Intelligent Systems Laboratory, Department of Chemical Engineering, Columbia University, New York City, NY 10027, U.S.A.}

\author{Manasi Khandekar} 
\affiliation{Department of Electrical Engineering, Columbia University, New York, NY 10027, U.S.A}

\author{Abhishek Sivaram}
\affiliation{Department of Chemical and Biochemical Engineering, Technical University of Denmark, 2800 Kongens Lyngby, Denmark}

\author{Collin Szczepanski} 
\affiliation{Complex Resilient Intelligent Systems Laboratory, Department of Chemical Engineering, Columbia University, New York, NY 10027, U.S.A.}







\maketitle

\section*{abstract}
Despite the stunning progress recently in large-scale deep neural network applications, our understanding of their microstructure, 'energy' functions, and optimal design remains incomplete. Here, we present a new game-theoretic framework, called statistical teleodynamics, that reveals important insights into these key properties. The optimally robust design of such networks inherently involves computational benefit-cost trade-offs that are not adequately captured by physics-inspired models. These trade-offs occur as neurons and connections compete to increase their effective utilities under resource constraints during training. In a fully trained network, this results in a state of arbitrage equilibrium, where all neurons in a given layer have the same effective utility, and all connections to a given layer have the same effective utility. The equilibrium is characterized by the emergence of two lognormal distributions of connection weights and neuronal output as the universal microstructure of large deep neural networks. We call such a network the Jaynes Machine.  Our theoretical predictions are shown to be supported by empirical data from seven large-scale deep neural networks. We also show that the Hopfield network and the Boltzmann Machine are the same special case of the Jaynes Machine.



\section{Statistical teleodynamics, Potential games, and Arbitrage equilibrium}\label{Statistical teleodynamics, Population games, and Arbitrage equilibrium}

Over forty years, the prevailing theory of neural networks has been the physics-based approach, where one defines an "energy" function that is minimized as the network learns the input-output patterns. Canonical examples of this approach are the Hopfield network~\cite{hopfield1982neural} and the Boltzmann machine~\cite{ackley1985boltzmann} and their variants~\cite{lecun2015deeplearning, bengio2017equil}. On the other hand, the practice of neural network training and deployment is dominated by the backpropagation algorithm combined with gradient descent. However, these dominant frameworks are unable to answer some fundamental questions about deep neural networks. \\

For example, given the obvious importance of connection weights in a neural network, what is the distribution of weights we ought to see in a fully trained network, and why?  We know what the "energy" of the network is, but what is its "entropy"? How is it related to the performance of the network? \\

The Hopfield model and the Boltzmann Machine model cannot answer these important questions. The backpropagation algorithm with gradient descent that does the actual job of training the network by tuning the weights cannot either. They are unable to answer because they lack an important perspective on this problem. This is what we address in this paper. \\

We present an alternative framework based on potential game theory that breaks new ground in our understanding of deep neural networks, revealing new insights about their microstructure. In a typical deep neural network training regimen using gradient descent, the backpropagation algorithm and regularization procedures nudge all weights and biases so that the overall error function is minimized over many iterations and over many datasets. Thus, the weights and biases are driven towards their optimal values iteratively. This movement of connections and neurons in the parameter space can be modeled equivalently as the iterative progress the connections and neurons make in a benefit-cost trade-off competition to improve their fitness or utility towards their optimal values. This naturally lends itself to a game-theoretic approach, which we formulate using a framework called~\textit{statistical teleodynamics}~\cite{venkat2004secon, venkat2006maxentropy, venkat2007darwin, venkat2015howmuch, venkat2017book, venkat2024active2}. It is a synthesis of the central concepts and techniques of population game theory with those of statistical mechanics. \\

In population games, one is interested in predicting the final outcome(s) of a large population of goal-driven agents competing dynamically to increase their respective utilities. In particular, one would like to know whether such a game would lead to an equilibrium outcome~\cite{easley2010networks, sandholm2010population}. For some population games, one can identify a single scalar-valued global function, called a {\em potential} $\phi(\boldsymbol{x})$ (where $\boldsymbol{x}$ is the state vector of the system) that captures the necessary information about the utilities of the agents. The gradient of the potential is the \textit{utility}. Such games are called {\em potential games}~\cite{rosenthal1973class,sandholm2010population, easley2010networks, monderer1996potential}. A potential game reaches strategic equilibrium, called \textit{Nash equilibrium}, when the potential $\phi(\boldsymbol{x})$ is maximized. Furthermore, this equilibrium is unique if $\phi(\boldsymbol{x})$ is strictly concave (i.e. $\partial^2 \phi /\partial^2 x < 0$)~\cite{sandholm2010population}. \\

Therefore, an agent's utility, $h_k$, in state $k$ is the gradient of potential $\phi(\boldsymbol{x})$, i.e.,
\begin{equation}
{h}_k(\boldsymbol{x})\equiv {\partial \phi(\boldsymbol{x})}/{\partial x_k}
\label{eq:utility}
\end{equation}
where $x_k=N_k/N$, $\boldsymbol{x}$ is the population vector, {$N_k$ is the number of agents in state $k$, and $N$ is the total number of agents}. By integration, we have 
\begin{eqnarray}
\phi(\boldsymbol{x})&=&\sum_{k=1}^m\int {h}_k(\boldsymbol{x}){d}x_k \label{eq:potential}
\end{eqnarray}

where $m$ is the total number of states. \\

To determine the maximum potential, one can use the method of Lagrange multipliers with $\mathscr{L}$ as Lagrangian and $\lambda$ as Lagrange multiplier for the constraint $\sum_{k=1}^mx_k=1$:
\begin{equation}
\mathscr{L}=\phi+\lambda(1-\sum_{k=1}^mx_k)
\label{eq:lagrangian}
\end{equation}

In equilibrium, all agents enjoy the same utility, that is, $h_k = h^*$. It is an \textit{arbitrage equilibrium} \cite{kanbur2020occupational} where agents no longer have any incentive to switch states, as all states provide the same utility $h^*$. Thus, the maximization of $\phi$ and $h_k = h^*$ are equivalent when the equilibrium is unique (i.e., $\phi(\boldsymbol{x})$ is strictly concave \cite{sandholm2010population}). \\

We use this formalism to model the competitive dynamics between neurons and between connections in a deep neural network, as we discuss next. \\

\section{Effective utility of  connections and the distribution of weights}\label{Effective utility of a connection and the distribution of weights}

During training, there are two \textit{local} competitions that occur simultaneously, one between the connections and the other between the neurons, in every layer. In these competitions, the agents (i.e., the connections and the neurons) move around in the parameter space, trying to minimize the error function in each iteration. We first consider the competition between the connections.\\

We consider a large deep neural network with $L$ layers of neurons. Let layer $l$ have $N^l$ neurons that are connected to neurons in layer $l-1$ using $M^l$ connections. To benefit from the statistical properties of large numbers, we assume that $N^l$ and $M^l$ are very large, on the order of millions. These connections have weights, which can be positive or negative, which determine the strength of the connections. In our analysis, we scaled the magnitude of all weights in the range of 0 to 1, which is divided into $m$ bins. Therefore, any given connection belongs to one of the $m$ bins. The strength of connection of a neuron $i$ in layer $l$ that is connected to a neuron $j$ in layer $l-1$, and belonging to bin $k$, is denoted by $w^l_{ijk}$. The number of connections in bin $k$ is given by $M^l_k$ with the constraint $M^l = \sum_{k=1}^{m} M^l_k$. The total budget for weights is constrained by $W^l = \sum_{k=1}^{m} M^l_k \mid w^l_{ijk} \mid$.\\

The deep neural network is human-engineered or has naturally evolved to meet certain information modeling goals and deliver certain performance targets efficiently and robustly\cite{venkat2004secon, venkat2006maxentropy, venkat2007darwin, venkat2011supplychain}. In our context, efficiency is a measure of how effectively the network minimizes the error or loss function with minimal use of resources. For example, maintaining neurons and connections incurs costs such as computing power, memory, time, energy, etc. One would like the network to meet its performance target of making accurate predictions with minimal use of such resources. Similarly, by robustness, we mean the ability to deliver the performance target despite variations in its operating environment, such as making accurate predictions in \textit{test} datasets that are different from its \textit{training} datasets.\\

We define the \textit{effective utility}, $h^l_{ijk}$, of a connection with a weight of $w^l_{ijk}$ in the layer $l$ as a measure of the contribution that this connection makes to reducing the error function robustly. The connections try to improve $h^l_{ijk}$ in each iteration to minimize the error function. The effective utility of a connection is a benefit-cost trade-off function. 
It is the net benefit contributed by a connection after accounting for the costs of maintenance and competition, as we discuss below. We assume that this competition is local, i.e., the connections $w^l_{ijk}$ compete only with each other for layer $l$ and not with the connections for the other layers. \\

Thus, the effective utility $h^l_{ijk}$ is made up of three components, 
\begin{equation}
     h^l_{ijk}  = u^l_{ijk} - v^l_{ijk} - s^l_{ijk}
     \label{eq:utility-uvs}
\end{equation}

where $u^l_{ijk}$ is the informational benefit derived from the strength of the connection, $v^l_{ijk}$ is the cost or disutility of maintaining such a connection, and $s^l_{ijk}$ is the disutility of competition between the connections. Disutilities are costs to be subtracted from the benefit $u^l_{ijk}$ to determine the net benefit.  \\

Now, in general, as the strength of the connection $w^l_{ijk}$ increases, the marginal utility of its contribution decreases. This diminishing marginal utility is a common occurrence for many resources and is usually modeled as a logarithmic function~\cite{venkat2017book, venkat2024active2}. Therefore, the utility $u^l_{ijk}$ can be written as

\begin{equation}
     u^l_{ijk}  = \alpha \ln \mid w^l_{ijk}\mid
     \label{eq:utility-u}
\end{equation}
where $\mid w^l_{ijk} \mid$ signifies that  $u^l_{ijk}$ depends only on the absolute magnitude and not on the sign of the weight, and $\alpha > 0$ is a parameter. \\

But, as noted, this benefit comes with a cost, as building and maintaining connections are not free. In biology, there are metabolic and energetic costs in creating and maintaining molecules and reactions associated with connections. In artificial neural networks, there are computational and performance costs. For example, it is well known that as weights increase, key performance metrics such as generalization accuracy and training time suffer, necessitating various regularization techniques. Such costs are taken into account in $v^l_{ijk}$. As Venkatasubramanian has shown, the appropriate model for this cost is a quadratic function, which has been successfully demonstrated for other dynamical systems, such as the emergence of income distribution~\cite{venkat2017book}, flocking of birds~\cite{venkat2022garuds}, and social segregation~\cite{venkat2024social}. \\

Therefore, we have $v^l_{ijk} = \beta (\ln \mid w^l_{ijk} \mid)^2$, such that
\begin{equation}
     u^l_{ijk} - v^l_{ijk}  = \alpha \ln \mid w^l_{ijk} \mid - \beta (\ln \mid w^l_{ijk} \mid) ^2
     \label{eq:utility-u-v}
\end{equation} where $\beta > 0$ is another parameter. \\

As more and more connections accumulate in the same bin $q$ (that is, having the same weight), each new connection is less valuable to the neuron. In other words, a connection is less valuable if it is one of the many rather than one of the few in its class. Therefore, a neuron would prefer the connections to be distributed over all the bins and not have them concentrated in just a few bins. This is enforced by another cost term $s^l_{ijk}$, the  \textit{competition cost}. Appealing to diminishing marginal (dis)utility again, we model this as $\gamma \ln M^l_k$, where $\gamma > 0$ is another parameter. This choice has also been successfully demonstrated for other systems, such as in the emergence of income distribution~\cite{venkat2017book}, flocking of birds~\cite{venkat2022garuds}, ant crater formation~\cite{venkat2022unified}, bacterial chemotaxis~\cite{venkat2024active2}, and social segregation~\cite{venkat2024social}.\\

 Therefore, the effective utility $h^l_{ijk}$ is given by 
\begin{equation*}
      h^l_{ijk}  = \alpha \ln \mid w^l_{ijk} \mid - \beta (\ln \mid w^l_{ijk} \mid) ^2 -\gamma \ln M^l_k
\end{equation*}

We can let $\gamma =1$ without any loss of generality and rewrite the equation as 

\begin{equation}
      h^l_{ijk}  = \alpha \ln \mid w^l_{ijk} \mid - \beta (\ln \mid w^l_{ijk} \mid) ^2 - \ln M^l_k
\label{eq:utility-alpha-beta-gamma1}
\end{equation}

All these connections compete with each other to increase their respective effective utilities ($h^l_{ijk}$) in their role to robustly reduce the overall error function. They do this by switching from one state to another by dynamically changing the weights $w^l_{ijk}$, depending on the local gradient of $h^l_{ijk}$, as in gradient descent. \\

One of the important results in potential game theory is that this competitive dynamics will result in a Nash equilibrium where the potential $\phi^l_w(x)$ is maximized. All agents enjoy the same utility in equilibrium – i.e., $h^l_{ijk} = h^{l*}$ in Eq.~\ref{eq:utility-alpha-beta-gamma1} for all $i, j$ and $k$, where the superscript ($*$) denotes the equilibrium state.  This is an \textit{arbitrage equilibrium} as all agents have the same utility, thus removing any incentive to switch states. \\

Using Eq. \ref{eq:utility-alpha-beta-gamma1} in Eq. \ref{eq:potential}, we have the layer connection-potential $\phi^l_w$ as 
 
\begin{equation}
\phi^l_w(\mathbf{x})=\phi^l_u + \phi^l_v + \phi^l_s + \text{constant}\label{pay_potential}
\end{equation}
where
\begin{eqnarray}
\phi^l_u &=& \alpha \sum_{k=1}^m x^l_k\ln \mid w^l_{ijk} \mid\\
\phi^l_v &=& -\beta \sum_{k=1}^m x^l_k(\ln \mid w^l_{ijk} \mid)^2\\
\phi^l_s &=& \frac{1}{M^l} \ln \frac{M^l!}{\prod_{k=1}^m(M^lx_k)!}
\label{eq:pot-connections}
\end{eqnarray}

where $x_k=M^l_k/M^l$ and we have used Stirling's approximation in Eq.~\ref{eq:pot-connections}.\\

We see that $\phi^l_w(\mathbf{x})$ is strictly concave:
\begin{eqnarray}
{\partial^2 \phi^l_w(\mathbf{x})}/{\partial x_k^{2}}=-{1}/{x_k}<0
\end{eqnarray}

Therefore, a {\em unique Nash Equilibrium} for this game exists, where $\phi^l_w(\mathbf{x})$ is maximized. Using the Lagrangian multiplier approach (Eq.~\ref{eq:lagrangian}), we maximize $\phi^l_w(\mathbf{x})$ in Eq.~\ref{pay_potential}-\ref{eq:pot-connections} to determine that the equilibrium distribution of the connection weights follows a \textit{lognormal distribution}, given by

\begin{equation}
x^*_k=\frac{1}{\sqrt{2\pi} \sigma_w \mid w^l_{ijk} \mid }\exp\left[-\frac{(\ln \mid w^l_{ijk} \mid - \mu_w)^2}{2\sigma_w^{2}} \right]
\label{eq:connection-lognormal}
\end{equation}\\

where, $\mu_w = \frac {\alpha + 1}{2\beta}$ and $\sigma_w = \sqrt{\frac{1}{2\beta}}$. \\

Thus, the theory predicts that for a \textit{fully trained} deep neural network, its microstructure, i.e. the distribution of connection weights, is lognormal for all layers. This universality is independent of its macroscopic architecture or its application domain. \\

The intuitive explanation is that, in a given layer, all individual connections contribute an effective utility (i.e., a net benefit) toward the overall objective of the network, which is to robustly minimize the error function.  In a large deep neural network, with hundreds of layers and millions of connections in each layer, no connection is particularly unique. No connection is more important than another. Every connection has thousands of counterparts elsewhere, so no one is special. Therefore, there is this inherent \textit{symmetry} and \textit{equality} in the microstructure. So, when training is completed, one reaches the arbitrage equilibrium where all effective utilities are equal in that layer, i.e., $h^l_{ijk} = h^{l*}$ for all $i,j$, and $k$.\\

Furthermore, in the “thermodynamic limit” of extremely large networks, i.e. $L \rightarrow \infty$, $M \rightarrow \infty$, and $W \rightarrow \infty$, \textit{all} connections in \textit{all} the layers end up making the \textit{same} effective utility contribution, i.e. $h^l_{ijk} = h^*$ for all $i,j,k$, and $l$. Therefore, all layers will have a lognormal weight distribution for this ideal deep neural network with the \textit{same} $\mu$ and $\sigma$. In other words, $\alpha$ and $\beta$ are the \textit{same} for all layers. This is the ultimate universal microstructure for ideal deep neural networks. We discuss the tests of these predictions with real-life deep neural networks in Section~\ref{Empirical Results}. \\

Now, readers familiar with statistical mechanics will recognize the potential component  $\phi^l_s$  as {\em entropy} (except for the missing Boltzmann constant $k_B$). Thus, by maximizing $\phi^l_w$ in the Lagrangian multiplier formulation, one is equivalently maximizing entropy subject to the constraints specified in the terms $\phi^l_u$ and $\phi^l_v$. Thus, the lognormal distribution is the maximum entropy distribution under these constraints. This connection with entropy reveals an important insight into the robustness property of network design, as discussed in Section~\ref{Optimally Robust Design}.\\

For the entire network of $L$ layers, we have the network connection-potential $\Phi_w$ as  
\begin{equation}
\Phi_w  = \sum_{l=1}^{L}\phi^l_w = \sum_{l=1}^{L}(\phi^l_u + \phi^l_v + \phi^l_s) 
\label{eq:POT-W-short}
\end{equation}
\begin{equation}
\Phi_w = \sum_{l=1}^{L} \sum_{k=1}^m \Big[\alpha x^l_k\ln \mid w^l_{ijk} \mid 
        - \beta x^l_k(\ln \mid w^l_{ijk} \mid)^2 \Big] + \mathscr{S}_{_{w}}
\label{eq:POT-W-long}
\end{equation}

where $\mathscr{S}_{_{w}} = \sum_{l=1}^{L}\phi^l_s$ is the network-wide connection-entropy of all connections in all layers. \\

Note that the error or loss function does not appear in our equations for the potential, as we are only considering the equilibrium state, which is at the end of the training of the network when the error is zero for the ideal network. \\

This ideal deep neural network is the conceptual equivalent of the ideal gas in statistical thermodynamics. Just as the maximum entropy distribution of energy in statistical thermodynamics is the well-known exponential distribution, called the Boltzmann distribution, we observe that its equivalent for connection weights in statistical teleodynamics is the lognormal distribution.\\

\section{Effective utility of neurons and the distribution of neuronal iota}\label{Effective utility of a neuron and the distribution of neuronal output}

Now that we have analyzed the network from the perspective of connections, let us consider the competition between neurons.  Most of the above analysis also applies to this perspective, with some changes reflecting the accommodations we need to make for neurons. \\

Here, the competition is between all neurons, $N^l$, in the layer $l$. As we did with the competition between connections, we assume that this competition also is \textit{local}, i.e., the neurons in a layer $l$ compete only with each other, and not with the neurons in the other layers. Each neuron performs two main tasks of information processing. The first is to compute the weighted sum of all the signals it receives from the neurons to which it is connected in the layer $l-1$.  For a neuron $i$ in layer $l$ that is connected to $N^{(l-1)}$ neurons in layer $l-1$, this is given by 

\begin{equation}
z^l_{i} = \sum_{j=1}^{N^{(l-1)}}w^l_{ij}y^{l-1}_j + b^l_i
\label{eq:neuron_input_z_i}
\end{equation}

where $y^{l-1}_j$ is the output of the neuron $j$ in layer $l-1$, $w^l_{ij}$ is the connection weight between the neurons $i$ and $j$, and $b^l_i$ is the bias of the neuron $i$ in layer $l$. The second task is to generate an appropriate output response $y^l_i$ from the input $z^l_i$ using its activation function and send it to all neurons to which it is connected in the layer $l+1$. Combining these two information-processing tasks, we formulate
\begin{equation}
    Z^l_i = z^l_i y^l_i = \left(\sum_{j=1}^{N^{(l-1)}} w^l_{ij} y^{l-1}_j + b^l_i\right) y^l_i
    \label{eq:Z_i}
\end{equation}

We believe that $Z^l_i$ is an important fundamental quantity of information processing that should be recognized with its own name and identity. We call this product of the input datum and the output datum, an \textit{iotum} (plural, \textit{iota}). It can be thought of as a quantum of information processing activity by the neuron $i$ in the layer $l$ towards the minimization of the error function. Like the connection weight, it is the neuron "weight." In the remainder of the paper, we assume a ReLU activation function, and, therefore, $Z^l_i > 0$. We compute the \textit{iota} $Z^l_i$ for all $N^l$ neurons in layer $l$, determine the minimum and maximum values, and divide the range into $n$ bins ($n \ll N^l)$. Therefore, each iotum will be in one of these bins, say, the $q^{th}$ bin of value $Z^l_q > 0$), and let the number of neurons in the $q^{th}$ bin be $N^l_q$. \\

Similar to our effective utility model of the weights in Eq.~\ref{eq:utility-uvs} and Eq.~\ref{eq:utility-alpha-beta-gamma1}, the effective utility $ H^l_q$ of a neuron in the $q^{th}$ bin for layer $l$ is given by
\begin{equation}
     H^l_q  = U^l_q - V^l_q - S^l_q
     \label{eq:utility-neuron-UVS}
\end{equation}

where $U^l_q$ is the informational benefit provided by the neuron in state $q$ by processing $Z^l_q$, and $V^l_q$ and $S^l_q$ are the computational and competition costs, respectively, incurred by the neuron in this activity. As in the case of weights (Eq.~\ref{eq:utility-alpha-beta-gamma1}), we have ($\eta > 0, \textcolor{red}{\zeta}  > 0$)
\begin{equation}
     U^l_q  = \eta \ln Z^l_q
     \label{eq:utility-neuron-U-lnZ}
\end{equation}
\begin{equation}
     V^l_q  = \zeta  (\ln Z^l_q)^2
     \label{eq:utility-neuron-V-lnZ}
\end{equation}
\begin{equation}
     S^l_q  = \ln N^l_q
     \label{eq:utility-neuron-S-lnN_q}
\end{equation}

and Eq.~\ref{eq:utility-neuron-UVS} becomes
\begin{equation}
     H^l_q  = \eta \ln Z^l_q - \zeta (\ln Z^l_q)^2 - \ln N^l_q
     \label{eq:utility-neuron-final}
\end{equation}

As we have shown for connections, this competition among neurons will also reach a unique arbitrage equilibrium where all neurons will have the same effective utility, $H^{l*}$. Therefore, we have
\begin{equation}
       H^{l*}  = \eta \ln Z^{l}_q - \zeta  (\ln Z^{l}_q)^2 - \ln N^{l*}_q
\label{eq:utility-neuron-equil}
\end{equation}

Again, as in Eq.~\ref{eq:connection-lognormal}, we get a \textit{lognormal} distribution 

\begin{equation}
x^{l*}_q=\frac{1}{\sqrt{2\pi}\sigma_{_N} Z^l_q }\exp\left[-\frac{(\ln Z^l_q - \mu_{_N})^2}{2\sigma_{_N}^{2}} \right]
\label{eq:neuron-lognormal}
\end{equation}\\

where, $x^{l*}_q=N^{l*}_q/N^l$, $\mu_{_N} = \frac {\eta + 1}{2\zeta}$, and $\sigma_{_N} = \sqrt{\frac{1}{2\zeta} }$. \\

From Eq.~\ref{eq:potential}, we can determine the layer neuron-potential $\phi^l_{_{N}}$ as 
\begin{eqnarray}
\phi^l_{_{N}} &=& \sum_{q} \int H^l_q d N^l_q \nonumber \\
            &=& \eta \sum_{q=1}^n x^l_q\ln Z^l_q
            -\zeta  \sum_{q=1}^n x^l_q(\ln Z^l_q)^2
            + \frac{1}{N^l} \ln \frac{N^l!}{\prod_{q=1}^n(N^l_q)!} \nonumber\\
            &=& \eta \sum_{q=1}^n x^l_q\ln Z^l_q
            -\zeta  \sum_{q=1}^n x^l_q(\ln Z^l_q)^2 + S^l_{_N} 
 \label{eq:pot-N-Z-q}
\end{eqnarray}

where $S^l_{_N} = \frac{1}{N^l} \ln \frac{N^l!}{\prod_{q=1}^n(N^l_q)!}$ is the layer neuron-entropy. \\

For the entire network of $L$ layers, we have the network-wide neuron-potential $\Phi_{_N}$ as  
\begin{equation}
\Phi_{_N} = \sum_{l=1}^{L}\phi^l_{_N} 
        =   \sum_{l=1}^{L} \sum_{q=1}^{n} \Big[\eta x^l_q\ln Z^l_q -
\zeta  x^l_q(\ln Z^l_q)^2 \Big] + \mathscr{S}_{_{N}}
\label{eq:POT-N-Z_q}
\end{equation}

where $\mathscr{S}_{_{N}} = \sum_{l=1}^{L} S^l_{_N}$ is the network-wide neuron-entropy. The neurons in a given layer $l$ reach the arbitrage equilibrium when $\phi^{l}_{_{N}}$ is maximized, and all neurons in all layers reach the arbitrage equilibrium when $\Phi_{_N}$ is maximized at the end of the training of the entire network.\\

Once again, note that the error or loss function does not appear in our equations for potential functions. We are only considering the equilibrium state, which is at the end of the network's training when the error is zero for the ideal network. \\

For the ReLU activation function, we have from Eq.~\ref{eq:neuron_input_z_i} 
\begin{eqnarray}
    y^l_i = \text{ReLU}(z^l_i) = \text{ReLU}\left(\sum_{j=1}^{N(l-1)} w^l_{ij} y^{(l-1)}_j + b^l_i\right)
    \label{eq:ReLU-y_i}
\end{eqnarray}

where $y^l_i$ is the ReLU activation output of the $i^{\text{th}}$ neuron in the $l^{\text{th}}$ layer. \\

Given that $y^l_i = \text{ReLU}(z^l_i) = 0$ for $z^l_i < 0$ and $y^l_i = z^l_i$ for $z^l_i > 0$,
\begin{equation}
    Z^l_q = z^l_q y^l_q = (z^l_q)^2 = (y^l_q)^2
    \label{eq:ReLU-Z-q}
\end{equation} 
and 
\begin{equation}
    Z^l_i = z^l_i y^l_i = (z^l_i)^2 = (y^l_i)^2
    \label{eq:ReLU-Z-i}
\end{equation}

Using Eq.~\ref{eq:ReLU-Z-q} in Eq.~\ref{eq:utility-neuron-equil}, we have
\begin{equation}
       H^{l*}  = \hat{\eta} \ln y^{l}_q - \hat{\zeta}  (\ln y^{l}_q)^2 - \ln N^{l*}_q
\label{eq:utility-neuron-relu}
\end{equation}

where $\hat{\eta} = 2\eta$ and $\hat{\zeta} = 4\zeta$. This leads to the lognormal distribution in $y^l_q$

\begin{equation}
x^{l*}_q = \frac{1}{\sqrt{2\pi}\hat{\sigma}_{_N} y^l_q }\exp\left[-\frac{(\ln y^l_q - \hat{\mu}{_{_N})^2}}{2\hat{\sigma}_{_N}^2} \right]
\label{eq:neuron-lognormal-relu-x}
\end{equation}

where, $x^{l*}_q=N^{l*}_q/N^l$, $\hat{\mu}_{_N} = \frac {\hat{\eta} + 1}{2\hat{\zeta} }$, and $\hat{\sigma}_{_N} = \sqrt{\frac{1}{2\hat{\zeta} }}$. \\

Therefore, for the special case of ReLU, the neuronal output $y^l_q$ follows a lognormal distribution. We tested this prediction with VGGNet-16, as discussed in Section~\ref{Empirical Results}. \\

In summary, during training, there are two local competitions that are happening simultaneously, one between the connections and the other between the neurons in every layer.  At the end of the training, both reach their own arbitrage equilibrium when $\Phi_w$ and $\Phi_{_N}$ are maximized, respectively. These arbitrage equilibria result in the emergence of two different lognormal distributions, one for connection weights and the other for neuronal iota. \\

\section{Optimally Robust Design}\label{Optimally Robust Design}

As we discussed in Sections~\ref{Effective utility of a connection and the distribution of weights} and \ref{Effective utility of a neuron and the distribution of neuronal output}, we see from Eq.~\ref{eq:POT-W-long} and Eq.~\ref{eq:POT-N-Z_q} that maximizing the potential ( $\Phi_w$ or $\Phi_{_N}$) is equivalent to maximizing the entropy ($\mathscr{S}_w$ or $\mathscr{S}_{_N}$) with the appropriate constraints. Hence, we call this network design procedure the \textit{maximum entropy design}~\cite{venkat2004secon, venkat2006maxentropy, venkat2007darwin, venkat2011supplychain}. The maximum entropy design distributes the connection weights (given by the lognormal distribution in Eq.~\ref{eq:connection-lognormal}) and neuronal iota (given by the lognormal distribution in Eq.~\ref{eq:neuron-lognormal} or Eq.~\ref{eq:neuron-lognormal-relu-x})  in the network in such a way that it maximizes the information-theoretic uncertainty about a wide variety of future datasets whose nature is unknown, unknowable, and, therefore, uncertain. Thus, in maximum-entropy design, the network is optimized for all potential future environments, not for any particular one. \\

Note that for any particular dataset, one can design a weight distribution such that it will outperform the maximum entropy design with respect to the error function. However, such a biased network may not perform as well for other datasets, while the maximum entropy distribution-based network is likely to perform better. For instance, if a network is overfitted on a specific dataset, then it might "memorize" these data and hence might not generalize that well for other datasets. \\

To prevent this, techniques such as data segmentation, weight regularization, dropout, early stopping, etc., are used~\cite{hinton2010Boltzmann, srivastava2014dropout}. The effect of such procedures is to achieve robustness in performance on a wide range of datasets. The goal of such techniques is to accommodate as much variability and as much uncertainty as possible in test environments. \\

This is what we achieve by maximizing entropy in our theory. Maximizing entropy is the same as maximizing the uncertainty and variability of future datasets. In our theory, this robustness requirement is naturally built in from the very beginning as an integral part of the effective utility and potential function formulation, not as \textit{ad hoc} afterthoughts to prevent overfitting. Therefore, the maximum entropy design leads to \textit{optimally robust design}. This design concept and the idea of using dropout to improve network performance were introduced by Venkatasubramanian earlier~\cite{venkat2004secon, venkat2007darwin, venkat2011supplychain} in a broader context of network design, which we have adapted here for deep neural networks.\\

Thus, an optimally robust deep neural network is a robust learning and prediction engine. It is a maximum entropy machine that learns an efficient and robust model of the target manifold. We call this machine the \textit{Jaynes Machine} in honor of Professor E. T. Jaynes, who elucidated the modern interpretation of the maximum entropy principle in the 1950s~\cite{jaynes1957information, jaynes1957information2, jaynes1979standonmaxentropy, jaynes1985wheredowego}. \\

\section{Empirical Results}\label{Empirical Results}

The predictions of the theory were tested by analyzing the distributions of connections and neurons in different networks. We first present the connection perspective, followed by the neuronal perspective. \\

\subsection{Distribution of connection weights}
For the distribution of connection weights, we tested using seven different deep neural networks. They are (i) BlazePose, (ii) Xception, (iii) VGGNet-16, (iv) BERT-Small, (v) BERT-Large, (vi) Llama-2 (7B), and (vii) LLAMA-2 (13B) \cite{BlazePose2023active, Xception2023active, BERT2023active, LLAMA2023active}. Their salient features are summarized in Table~\ref{tab:table1}. The first three utilize convolution layers, and the other four are based on the transformer architecture~\cite{LLAMA2web,BERTweb,BlazePoseweb}. They are of widely different sizes with respect to the number of parameters and are designed for different application domains. \\

The layer-by-layer weight data for these networks were extracted, normalized between 0 and 1, converted to their absolute magnitudes by dropping the signs, and classified into different bins. For all these networks, some layers had only a few thousand data points (out of the millions or tens of millions in the network), so we did not fit a distribution as statistical measures such as $R^2$ were not good. \\

\begin{table}
    \centering
    \begin{tabular}{c c c c}
    \hline
    Model & Architecture & Parameters size & Application \\ \hline
        BlazePose & Convolution & $2.8\times10^6$ & Computer Vision \\
        Xception & Convolution & $20\times10^6$ & Computer Vision \\
        VGGNet-16  & Convolution & $138\times10^6$ & Computer vision \\
        BERT Small & Transformer & $109\times10^6$ & NLP\\
        BERT Large & Transformer & $325\times10^6$ & NLP \\
        LLAMA-2 (7B) & Transformer & $7\times10^9$ & NLP\\
        LLAMA-2 (13B) & Transformer & $13\times10^9$ & NLP \\
    \hline
    \end{tabular}
    \caption{Seven deep neural network case studies}
    \label{tab:table1}
\end{table}

The plots show the \textit{size-weighted} distributions (noted as size-weighted count on the y-axis) rather than the weight distribution, since the features are clearer in the former. The size-weighted count of a bin is simply the product of the weight of the bin and the number of connections in that bin. There is a well-known result in statistics \cite{rao1984size} that if a variate is distributed lognormally with $\mu$ and $\sigma$ (i.e., $LN(\mu, \sigma)$), then the size-weighted distribution of the variate is also lognormal, $LN(\mu^{\prime}, \sigma^{\prime})$, where $\mu^{\prime} = \mu + \sigma^2$ and $\sigma^{\prime} = \sigma$. Furthermore, since the utility $u^l_{ijk}$ in Eq.~\ref{eq:utility-u} is positive (since it is a benefit) and $\ln \mid w^l_{ijk} \mid$ is negative in the range of $0 < \mid w^l_{ijk} \mid < 1$, we have $\alpha < 0$, $\mu < 0$, and $\mu^{\prime}< 0$. Similarly, from Eq.~\ref{eq:utility-u-v}, the disutility $v^l_{ijk}$ requires $\beta > 0$. The parameter $A^{\prime}$ is the scaling factor of the lognormal distributions.\\ 

\begin{figure}[!ht]
    \centering
    \includegraphics[width = 01\linewidth]{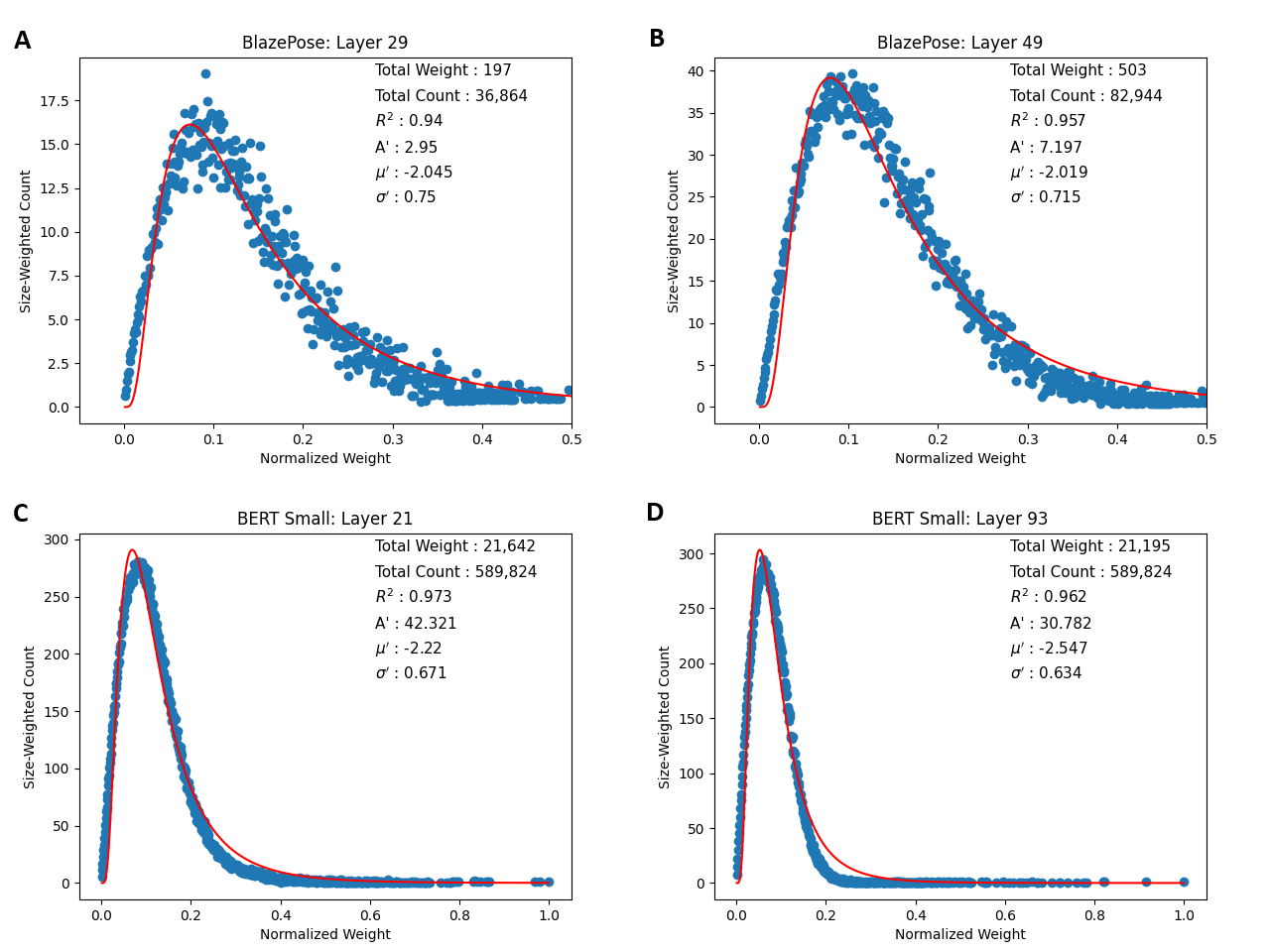}
    \caption{Typical lognormal fitted curves: A \& B - BlazePose; C \& D  - Xception. Blue dots are data, and the red curves are lognormal fits. The parameters of the fits are also shown.}
    \label{fig:BlazePose-Xception-lognormal}
\end{figure}
\begin{figure}[!ht]
    \centering
    \includegraphics[width = 01\linewidth]{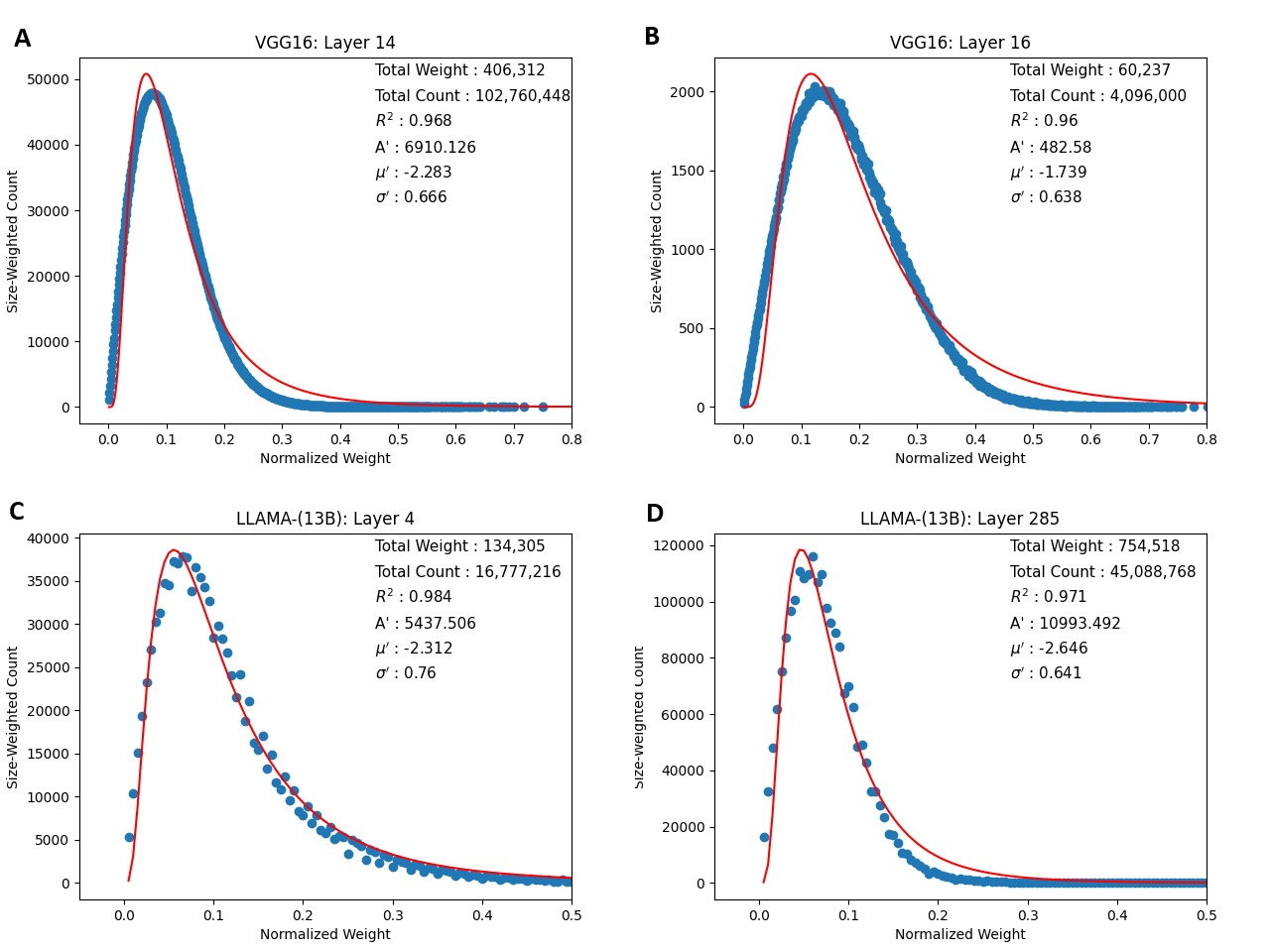}
    \caption{(A) \& (B): VGG16 Layer \#14 \& Layer \#16,(C) \& (D): LLAMA-2 7B Layer \#4 \& Layer \#285}
    \label{fig:VGG16-LLAMA-7B-lognormal}
\end{figure}

The lognormal distribution was fitted and tested for the following networks: BlazePose (39 layers), Xception (32 layers), VGGNet-16 (16 layers), BERT-Small (75 layers), BERT-Large (144 layers), Llama-2 7B (226 layers), and Llama-2 13B (282 layers). Instead of showing the plots for all the 814 layers, which all look pretty similar, we show a much smaller selection of sample distributions in Figs. \ref{fig:BlazePose-Xception-lognormal}-\ref{fig:VGG16-LLAMA-7B-lognormal}. We see that the size-weighted data fit the lognormal distribution very well with high $R^2$ values.  This is typical for all layers with high connectivity (see the Tables in the SI). Although the seven networks use different architectures, are of different sizes, and are trained for different applications, we find this surprising universal microstructure. This is an important design feature of these networks that emerges automatically during training. As discussed in Section~\ref{Effective utility of a connection and the distribution of weights}, our theory predicts this universal lognormal microstructure. \\

\begin{table}
    \centering
    \caption{Lognormal parameters for the size-weighted distribution of weights}
    \label{tab:table2}
    \begin{tabular}{c c c c c c}
    \hline
    Model & Layers & $R^2$ & $\mu^{\prime}$ & $\sigma^{\prime}$ \\ \hline
        BlazePose & 39 & $0.93 \pm 0.02$  & $-1.74 \pm 0.52 $ & $1.49 \pm 0.60 $\\
        Xception & 32 & $0.98 \pm 0.01$  & $-2.87 \pm 0.18 $ & $0.70 \pm 0.05 $\\
        VGGNet-16 & 16 & $0.97 \pm 0.01$ & $-2.36 \pm 0.41$ & $0.68 \pm 0.05$\\
        BERT Small & 75 & $0.96 \pm 0.01$  & $-2.47 \pm 0.95 $ & $0.65 \pm 0.02 $\\
        BERT Large & 144 & $0.96 \pm 0.01$  & $-2.37 \pm 0.98 $ & $0.64 \pm 0.01 $\\
        LLAMA-2 (7B) & 226 & $0.97 \pm 0.01$  & $-2.96 \pm 0.54 $ & $0.66 \pm 0.05 $\\
        LLAMA-2 (13B) & 282 & $0.94 \pm 0.03$  & $-3.02 \pm 0.53 $ & $0.67 \pm 0.06$ \\
    \hline
    \end{tabular}
    
\end{table}

In the Supplementary Information section, we list the lognormal parameters ($A^{\prime}$, $\mu^{\prime}$, and $\sigma^{\prime}$) for all highly connected layers for the seven case studies. Table~\ref{tab:table2} summarizes $\mu^{\prime}$ and $\sigma^{\prime}$ for the seven case studies. Note that for large networks with $>100$ million connections,  $\sigma^{\prime}$ appears to be nearly constant (around 0.65) for all networks, as seen by its low standard deviation values in Table~\ref{tab:table2}. This implies that $\beta^{\prime}$ is also approximately constant for all networks. Even $\mu^{\prime}$ (and hence $\alpha^{\prime}$) appears to be in a narrow range (-2.3 to -3.0) for the different networks. The theory predicts that $\mu^{\prime}$ and $\sigma^{\prime}$ are constants for all networks only in the "thermodynamic limit" of the ideal network. However, we see this trend even for these nonideal cases.\\

\begin{figure}[!ht]
    \begin{subfigure}{\linewidth}
        \centering
        \includegraphics[width=\linewidth]{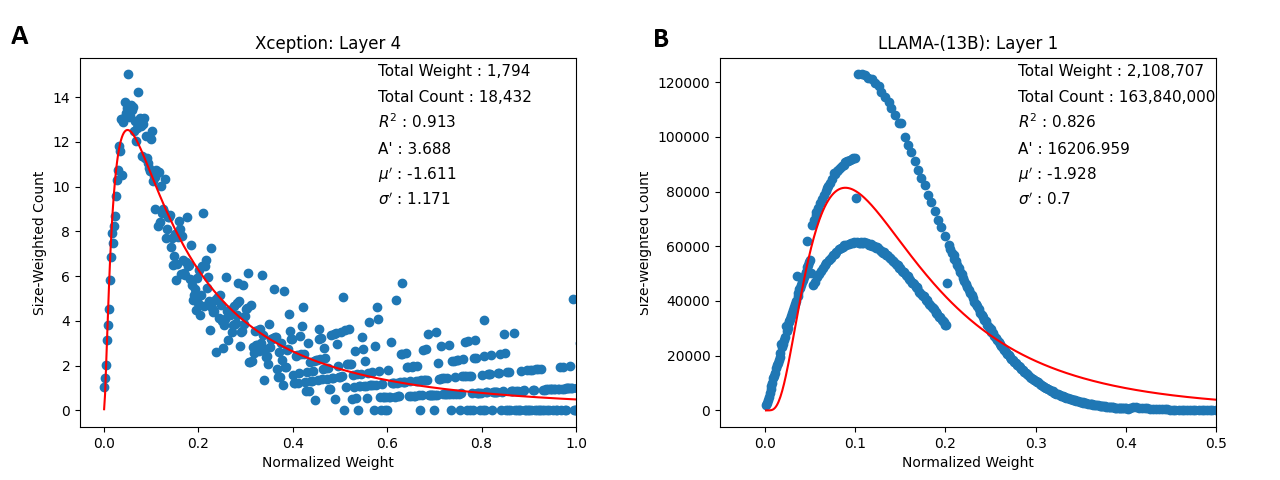}
    \end{subfigure}
    \caption{(A) Noisy data in a layer with low number of connections: Xception Layer \#4 (B) Suboptimal training in a highly connected layer: LLAMA-2 13B Layer \#1}
    \label{fig:noisy-suboptimal}
\end{figure}

The number of connections in the 814 layers we studied ranged from 36,864 to  163,840,000. Generally speaking, we find that the more connections a layer has, the better the lognormal fit with higher $R^2$ due to better statistical averaging. We can see in Fig.~\ref{fig:noisy-suboptimal}A that Layer \#4 of the Xception network, which has only 18,432 connections, has too much noise in the data to fit any distribution well. Therefore, we did not model such layers.\\

However, layers with scores of millions of connections have their own challenges, as they are harder to train, and hence run the risk of suboptimal weight assignments. Recall that, according to the theory, the lognormal distribution emerges only when the arbitrage equilibrium is reached. It is possible that these extremely highly connected layers had not quite reached equilibrium when the training was stopped. Therefore, there would be a mismatch between theoretical predictions and empirical observations. We observe this in the Llama-2 (13B) data. Fig. \ref{fig:noisy-suboptimal}B shows the size-weighted distribution for Layer \#1, which has over 163 million weights. As we can clearly see, there are elements of the lognormal distribution present, but the fit is not as good as it is for the smaller LLAMA-2 in Fig.~\ref{fig:VGG16-LLAMA-7B-lognormal}C-D. This suggests that Layer \#1 training was suboptimal. It appears from our empirical analysis that layers that have connections in the range of about 1 to 70 million have the right trade-off between better statistical properties and reaching optimal weight distribution.   \\

\subsection{Distribution of neuronal iota}

For the neuronal perspective, we tested the theory's prediction that the neuronal iota $Z^l_q$ in a given layer will follow a lognormal distribution. For ReLU-based networks, this implies that the neuronal output $y^l_q$  follows lognormal. We report the results for VGGNet-16, as we did not have access to the other networks to do these tests (Blazepose was not considered because it is too small for this test). \\

\begin{figure}[!ht]
    \begin{subfigure}{\linewidth}
        \centering
        \includegraphics[width=\linewidth]{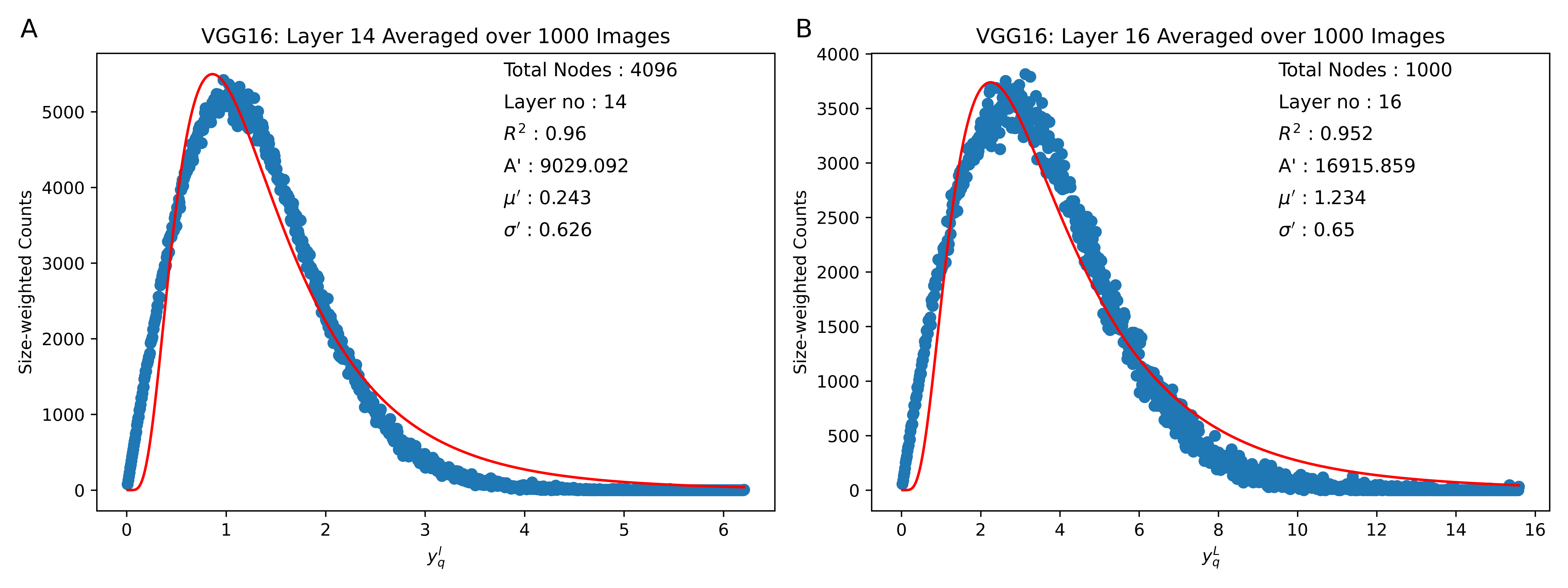}
    \end{subfigure}
    \caption{VGGNet-16: Lognormal distribution of size-weighted $y^l_q$ for (A) layer \#14 (B) layer \#16}
    \label{fig:VGG16-lognormal-14-16}
\end{figure}

We presented 1000 different images to VGGNet-16 and recorded the corresponding values of $y_i$ for all neurons in layers \#14-16 for each image. The values were classified into 1000 bins and the average value of $N^l_q$ over 1000 images was calculated for all bins. We then plotted the size-weighted counts as shown in Fig.~\ref{fig:VGG16-lognormal-14-16}. As predicted by the theory, we find that the lognormal distribution fits the data well. \\  

\section{Comparison with the Hopfield network and the Boltzmann machine}\label{Comparison with the Hopfield network and the Boltzmann machine}

It is instructive to compare the Jaynes Machine with the Hopfield network~\cite{hopfield1982neural}, and the Boltzmann Machine~\cite{ackley1985boltzmann}. \\

Consider a neuron's computational benefit, given by Eq.~\ref{eq:utility-neuron-U-lnZ}, and its computational cost, given by Eq.~\ref{eq:utility-neuron-V-lnZ}. Instead of these expressions, let them have the following simpler forms:
\begin{equation}
     U^l_q  = B^l
     \label{eq:utility-neuron-U-B}
\end{equation}
\begin{equation}
     V^l_q  = \zeta  Z^l_q
     \label{eq:utility-neuron-V-Z}
\end{equation}

where $B^l \geq 0$ and $\zeta  > 0$ are constants. Under these conditions, Eq.~\ref{eq:utility-neuron-equil} becomes
\begin{equation}
      H^{l*}  = B^l - \zeta  Z_q^{l} - \ln N_q^{l*}
\label{eq:utility-neuron-equil-hopfield}
\end{equation}
and we have 
\begin{equation}
    N_q^{l*}  =  e^{(B^l - H^{l*})} e^{-\zeta}  Z^l_q
    \label{eq:N_q-hopfield}
\end{equation}
which is an exponential distribution of the iota. \\

Furthermore, the layer neuron-potential becomes (ignoring the constant terms)
\begin{eqnarray}
\phi^l_{_{N}} &=& -\zeta  \sum_{i=1}^{N^l} Z^l_i + S^l_{_N}
\label{eq:pot-N-exponential}
\end{eqnarray}

and the network neuron-potential $\Phi_{_N}$ is  
\begin{eqnarray}
\Phi_{_N} &=& -\zeta   \sum_{l=1}^{L} \sum_{i=1}^{N^l} Z^l_i + \mathscr{S}_{_{N}} 
\label{eq:POT-N-exponential}
\end{eqnarray}

where $\mathscr{S}_{_{N}} = \sum_{l=1}^{L} S^l_{_N}$ is the network neuron-entropy. 
Now, maximizing the potential $\Phi_{_N}$ at arbitrage equilibrium implies minimizing $\sum_{l=1}^{L} \sum_{i=1}^{N^l} Z^l_i$. Therefore, from the definition of $Z^l_i$ in Eq.~\ref{eq:Z_i}, we get at equilibrium 
\begin{equation}
\text{Min} \sum_{l=1}^{L} \sum_{i=1}^{N^{l}}Z^l_{i} = \text{Min} \sum_{l=1}^{L} \sum_{i=1}^{N^{l}} \left( \sum_{j=1}^{N^{(l-1)}}w^l_{ij}y^{l-1}_j y^l_i +  b^l_i y^l_i \right)
\label{eq:POT-min-Hopfield}
\end{equation}

We recognize Eq.~\ref{eq:POT-min-Hopfield} as the "energy" function that is minimized at statistical equilibrium for the Hopfield network~\cite{hopfield1982neural} and the Boltzmann Machine~\cite{ackley1985boltzmann}. Thus, we see that these are the same special cases of the Jaynes Machine when the informational benefit is constant (i.e., $U^l_q = B^l$), and the computational cost $V^l_q$ is linear in $Z^l_q$. However, the Hopfield and Boltzmann models are incomplete, as they do not include the neuron-entropy  $\mathscr{S}_{_{N}}$ term. \\

Furthermore, the Hopfield network and the Boltzmann Machine imply an \textit{exponential} distribution of $Z^l_q$ at equilibrium (Eq.~\ref{eq:N_q-hopfield}), but the empirical evidence we presented above for the seven networks validates our theory's prediction that it is lognormal. In addition, neither the Hopfield model nor the Boltzmann Machine model can predict the distribution of weights, which our theory does correctly. \\

\section{Discussion and Conclusions}
In physics, we have systems that contain an Avogadro number ($\sim10^{23}$) of atoms or molecules that dynamically interact to produce a wide variety of macroscopic phenomena. The theory of these interactions is described by statistical mechanics in quantitative detail. Similarly, a large deep neural network is a dynamical system comprising millions or billions of interacting neurons and their connections. So, what is the corresponding theory of neural networks? By theory, we mean a fundamental organizing principle(s) that can predict important properties, such as the distribution of neuronal iota and connection weights in a well-trained network.  \\

The best candidate we have so far is the four-decade-old Hopfield network model, which is inspired by the Ising model in statistical mechanics. However, it is not a correct theory as its prediction of the distribution of neuronal iota does not agree with empirical evidence, as discussed above. Furthermore, it cannot predict the distribution of connection weights. \\

On the other hand, the enormously successful workhorse of neural network training, gradient descent combined with the backpropagation algorithm, is not a theory of neural networks. It cannot make any predictions about the distributions of neuronal iota or connection weights. It is only a calculus recipe, the chain rule, for tuning the parameters to get the desired outcome.  In a similar vein, even the highly effective architectural and algorithmic innovations, such as transformers and reinforcement learning, are also recipes to get things done. They do not offer a theory of neural networks. \\

In this paper, we address this need by proposing a new theory of learning in neural networks that considers a well-trained neural network as an \textit{optimally robust learning and prediction machine}. The machine's effectiveness is determined by its ability to make accurate predictions robustly under different conditions. Our theory formulates the learning process as an informational benefit-cost trade-off competition between neurons and between connections using a game-theoretic modeling framework called statistical teleodynamics. The theory predicts that such a competition will result in an arbitrage equilibrium, which is the final state of a fully trained network. At arbitrage equilibrium, the theory predicts that the connection weights and the neuronal iota are distributed lognormally. This microstructure is independent of the architecture or the application domain. We call this ideal network the Jaynes Machine. These predictions are supported by empirical evidence from artificial neural networks. Furthermore, recent studies with biological neural networks have also reported a lognormal weight distribution~\cite{ashhad2023lognormal, yuan2023lognormal}. \\

These results should help us develop custom training algorithms and special-purpose hardware to reduce the time, data, and computational resources required to train large deep neural networks effectively. \\

In our theory, the key concepts are the utility (benefit), the disutility (cost), and the effective utility (benefit-cost) of agents (neurons and connections), competition between agents to increase their effective utility, the game potential (fairness in the assignment of effective utility~\cite{venkat2017book}) of the entire network, and arbitrage equilibrium. These model the information processing activities of neurons and connections in the network towards the minimization of the error or loss function. \\

We identify \textit{iotum} ($Z^l_i$) as an important fundamental quantity of information processing performed by a neuron to minimize the error function. The informational benefits and computational costs of neurons depend on this key quantity. It is similar to the interaction energy in physicochemical systems. However, unlike energy, it is not conserved. 
Furthermore, in physicochemical systems, there is no notion of a "benefit" or a "cost" for the agents (i.e., atoms and molecules). Here, the equivalent of "effective utility" of a molecular agent (which is the chemical potential, see~\cite{venkat2015howmuch, venkat2017book} for more details) is given by (from Eq.~\ref{eq:utility-neuron-equil-hopfield}, $B^l = 0$, $\zeta  = \beta = 1/kT$)
\begin{equation}
      H^*  = - \beta E_q - \ln N_q^{*}
\label{eq:utility-energy}
\end{equation}

where $E_q$ is the energy of a molecule in state $q$, $\beta$ is the Boltzmann factor. This equation leads to the well-known Boltzmann exponential energy distribution at thermodynamic equilibrium. \\

In contrast, a neural network is a learning engine that has been optimally designed or evolved to make accurate predictions robustly under resource constraints. Therefore, its microstructure reflects the benefit-cost trade-offs made in its optimally robust design. Neither the Hopfield model nor the Boltzmann machine model captures this benefit-cost trade-off feature, as they originate from a physics perspective, which lacks these concepts. In this sense, our approach is philosophically different. By taking a game-theoretic perspective, we account for the benefit-cost trade-offs as modeled in Eq.~\ref{eq:utility-neuron-final} and Eq.~\ref{eq:POT-N-Z_q}. Because of this critical difference between a system that is designed/evolved for a purpose vs. one without such a purpose, we find the corresponding models (e.g., Eq.~\ref{eq:utility-neuron-final} vs. Eq.~\ref{eq:utility-energy}) to be quite different. \\

This comparison also highlights the meaning of potential $\Phi_{_N}$ and effective utility $H^l_q$ in the context of thermodynamics. As Venkatasubramanian has discussed at length \cite{venkat2015howmuch, venkat2017book, venkat2022unified}, $\Phi_{_N}$ is equivalent to the Helmholtz free energy (i.e., the Gibbs free energy ($G$) with $PV = 0$, where $P$ is pressure and $V$ is volume) with an important difference that $\Phi_{_N}$ is maximized, while the Gibbs free energy is minimized. Mathematically, Eq.~\ref{eq:POT-N-Z_q} (or Eq.~\ref{eq:POT-W-long}) in statistical teleodynamics for neural networks is equivalent to $G = E + PV -TS$ (with $PV = 0$) in statistical thermodynamics for physicochemical systems. One maximizes the potential in statistical teleodynamics to reach arbitrage equilibrium, whereas one minimizes the free energy in statistical thermodynamics to reach thermodynamic equilibrium. Just as minimizing the free energy is the organizing principle for passive matter systems, we suggest that maximizing the potential could be an important organizing principle for active matter systems, such as neural networks. \\

Similarly, the effective utility is equivalent to the chemical potential~\cite{venkat2015howmuch, venkat2017book, venkat2022unified}, with the important difference that the former is maximized, while the latter is minimized. Thermodynamic equilibrium is a special case of arbitrage equilibrium when the arbitrage currency is just the chemical potential. Although active matter systems such as neural networks are generally characterized as nonequilibrium or out-of-equilibrium systems~\cite{marchetti2013hydrodynamics}, our theory recognizes them as systems in arbitrage equilibrium. \\

Given this background, it is easy to see that the Hopfield and Boltzmann machine models are incomplete as they do not include the entropy term ($TS$). That is, instead of $G$, they only have the "energy" $E$, which is minimized.  As a result, \textit{ad hoc} regularization techniques, such as dropout, have to be imposed separately to make learning robust. Our potential functions, on the other hand, include the entropy term (whether $\mathscr{S}_{_{w}}$ or $\mathscr{S}_{_{N}}$) explicitly in Eq.~\ref{eq:POT-W-long} and Eq.~\ref{eq:POT-N-Z_q}. This built-in entropy term takes care of regularization right from the beginning.\\

We continue the comparison with thermodynamics a little more to observe that both Eq.~\ref{eq:POT-W-long} and Eq.~\ref{eq:POT-N-Z_q} each mathematically express the equivalent of the "First law" and the "Second law" of statistical teleodynamics. Maximizing potential $\Phi$ is the same as maximizing entropy ($\mathscr{S}_{_{w}}$ or $\mathscr{S}_{_{N}}$), which is the "Second law" component, under the constraints expressed by $\phi^l_u$ and $\phi^l_v$ for $\max \mathscr{S}_{_{w}}$ or by $\sum_{l=1}^{L} \sum_{q=1}^{n} \left( \eta x^l_q U^l_q - \textcolor{red}{\zeta}  x^l_q V^l_q \right) $ for $\max \mathscr{S}_{_{N}}$, which is the "First law" component. The "Zeroth law" of statistical teleodynamics is the postulate that all agents continuously strive to increase their individual effective utility by exploring all opportunities. This law captures the essence of the Darwinian survival-of-the-fittest principle in biology and the Smithian principle of the pursuit of self-interest in economics. \\

It is reassuring to know that the same mathematical framework has been successfully demonstrated for other dynamical systems, as summarized in Table \ref{tab:utility-domains}, in biology \cite{venkat2022unified}, ecology \cite{venkat2024active2}, sociology \cite{venkat2024social}, and economics \cite{venkat2015howmuch, venkat2017book, kanbur2020occupational} to predict emergent phenomena and arbitrage equilibria. \\

\begin{table}[!h]
    \centering
    \caption{Effective utility function in different domains}
    \label{tab:utility-domains}
    \scalebox{0.9}{\begin{tabular}{c c c }\\
    \hline
        \textbf{Domain}& \textbf{System}& \textbf{Utility function} ($h_i$) \\ \hline\\
         Physics & Thermodynamics & $-\beta  E_i - \ln n_i$ \\\\
         Biology & Bacterial chemotaxis & $\alpha c_i - \ln n_i$ \\\\
         Ecology & Ant crater formation & $b - \dfrac{\omega r_i^a}{a} - \ln n_i$ \\\\
         Ecology & Birds flocking & $\alpha n_i - \beta n_i^2+ \gamma n_i l_i - \ln n_i$ \\\\
         Sociology & Segregation dynamics & $\eta n_i - \xi n_i^2 + \ln(H-n_i) -\ln n_i$ \\\\
         Economics & Income distribution & $\alpha \ln S_i - \beta \left(\ln S_i\right)^2 - \ln n_i$ \\\\ 
         \hline
    \end{tabular}}
\end{table}

As noted, Eq.~\ref{eq:utility-energy} models the emergence of the Boltzmann distribution in thermodynamics. Similarly, in biological systems, the benefit-cost trade-off in effective utility $h_i$ for bacterial chemotaxis is modeled by~\cite{venkat2022unified}
\begin{equation}
h_{i}= \alpha c_i - \ln n_i
\end{equation}

where the first term is the benefit derived from a resource ($c_i$, $\alpha > 0$), and the second is the competition cost. In a similar vein, the emergence of ant craters is modeled by
\begin{equation}
    h_i = b - \frac{\omega r_i^a}{a} - \ln n_i
\end{equation}

where the first term ($b > 0$) is the benefit of having a nest for an ant, the second term is the cost of the work carried out by an ant to carry the sand grains to build the nest, and the last term is again the competition cost. The arbitrage equilibrium here results in the Weibull distribution~\cite{venkat2022unified}. Also, in ecology, the spontaneous emergence of flocking of bird-like agents has been demonstrated using the effective utility~\cite{venkat2022garuds}
\begin{equation}
    h_i = \alpha n_i - \beta n_i^2 + \gamma n_i l_i - \ln n_i \\
\end{equation}

where the first term is the benefit of the community of bird neighbors, the second is the congestion cost of such neighbors, the third is the benefit of flying in the same direction, and the last is again the competition cost.\\

In sociology, Schelling game-like segregation dynamics was demonstrated by the model~\cite{venkat2024social} 
\begin{eqnarray}
    h_i = \eta n_i - \xi n_i^2 + \ln(H - n_i) - \ln n_i
\end{eqnarray}

where the first term is the benefit of the community of neighbors, the second is the congestion cost of such neighbors, the third is the benefit of relocation options, and the last is again the competition cost. \\

Finally, in economics, the emergence of a lognormal income distribution is modeled by~\cite{venkat2015howmuch, venkat2017book}
\begin{equation}
h_{i}= \alpha \ln S_i - \beta \left(\ln S_i\right)^2 - \ln n_i
\end{equation}

where the first term is the benefit of the salary ($S_i$), the second is the cost of work expended to earn this salary, and the last is again the competition cost. Note that the constant parameters $\alpha, \beta, \gamma, \eta, \zeta,$ and $\xi$ are different in different applications. \\

Comparing these effective utility functions with those of connections (Eq.~\ref{eq:utility-alpha-beta-gamma1}) and neurons (Eq.~\ref{eq:utility-neuron-final}), we observe a certain \textit{universality} in their structure in different domains. They are all based on benefit-cost trade-offs, but the actual nature of the benefits and costs depend on the details of the specific domain, as one would expect. Thus, we see that the same conceptual and mathematical framework, statistical teleodynamics, is able to predict and explain the emergence of various distributions via arbitrage equilibrium in dynamical systems in physics, biology, ecology, sociology, economics, and computer science. \\

In all these utility functions, the competition cost term (-$\ln n_i$) is the same, which leads to maximum entropy when expressed in the potential $\Phi$ form, as we see in Eq.~\ref{eq:POT-W-long} and Eq.~\ref{eq:POT-N-Z_q}. As noted, this term corresponds to the "Second law" component, while the other terms enforce the "First law" constraints. \\

The crucial feature of the maximum entropy design, expressed by the lognormal distribution at the arbitrage equilibrium, is that the effective utilities of all the connections are equal. Similarly, the arbitrage equilibrium of neurons shows that their effective utilities are also equal. These invariance-like properties reflect a deep sense of symmetry, harmony, and fairness in optimal network design. This is an elegant solution to the credit assignment problem that involves millions of connections and neurons. What we have presented is the van der Waals equation version of statistical teleodynamics. Much more remains to be done to incorporate the different practical considerations that are seen in large, real-world deep neural networks.

\section*{Methods}
In this work, we consider the trained machine learning models -- BlazePose, Xception, VGGNet-16, BERT Small, BERT Large, LLAMA-2 (7B) and LLAMA-2 (13 B), available as open source (see the references). We consider only fully connected layers in these models for the weight and iota distributions.\\


\section*{Acknowledgements}
VV would like to thank V. Govind Manian and Babji Srinivasan for their valuable feedback.


\section*{Author Contributions}

VV: Conceptualization, Theory, Methodology, Formal Analysis, Investigation, Supervision, Funding Acquisition, and Writing; NS: Software development and formal analysis for the BlazePose, VGGnet-16, BERT-Small, BERT-Large, Llama-2 (7B), and Llama-2 (13B) networks; MK: Software development and formal analysis for the Xception network; AS: Formal analysis; CS: Formal analysis of neuronal utility \\

The authors have no conflicts of interest to declare.

\bibliography{PNASJaynes.bib}

\newpage

\end{document}



\maketitle

In this section, we provide seven tables that display a layer-by-layer summary of the lognormal parameters for the seven networks. Also displayed are the lognormal plots for the BERT (Small and Large) and the LLAMA-2 (7B and 13B) networks. 

\begin{figure}[!h]
    \centering
    \includegraphics[width = 0.75\linewidth]{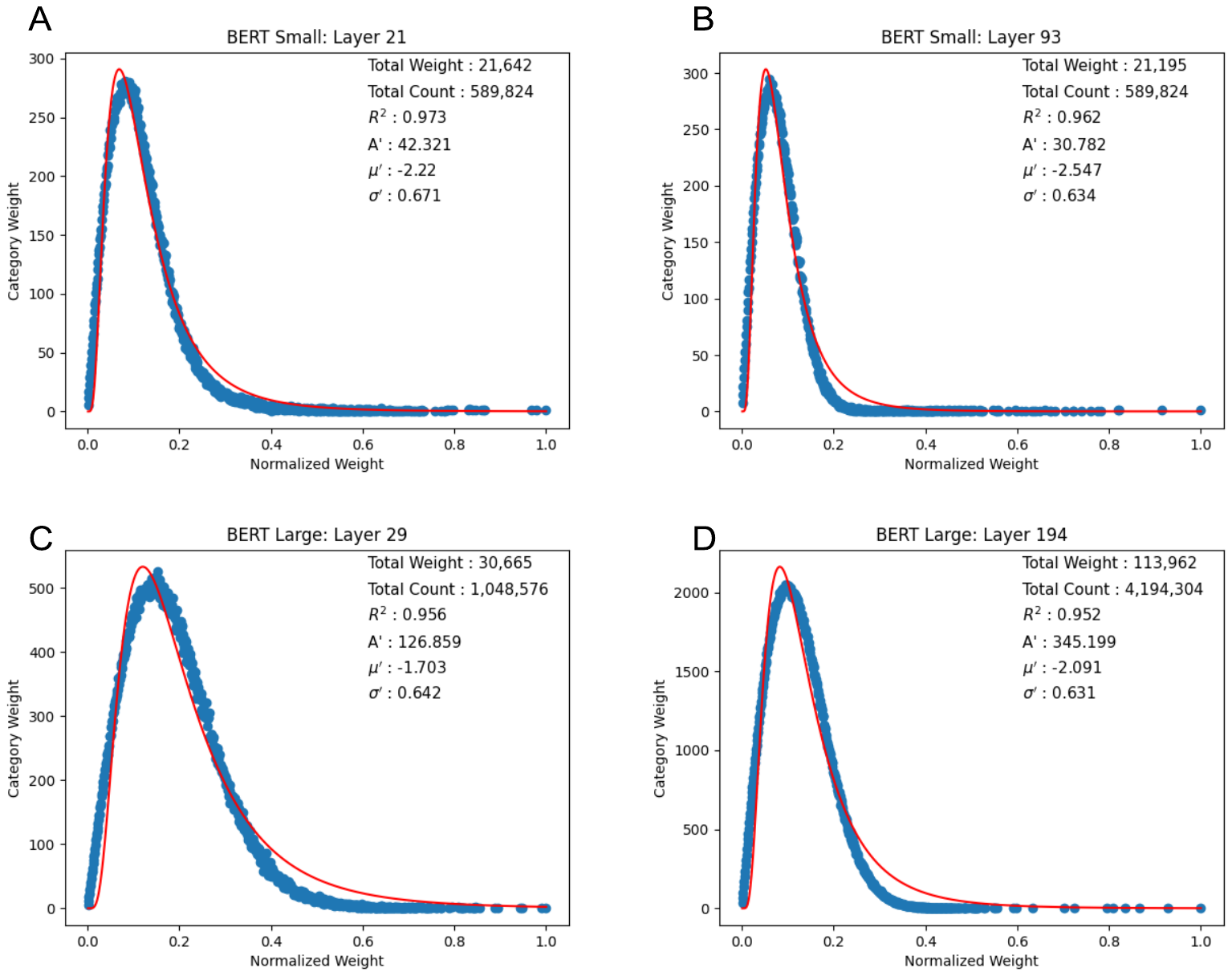}
    \caption{A \& B - BERT-Small; C \& D - BERT-Large}
    \label{fig:BERT}
\end{figure}

\newpage 

\begin{figure}[!h]
    \centering
    \includegraphics[width = 0.75\linewidth]{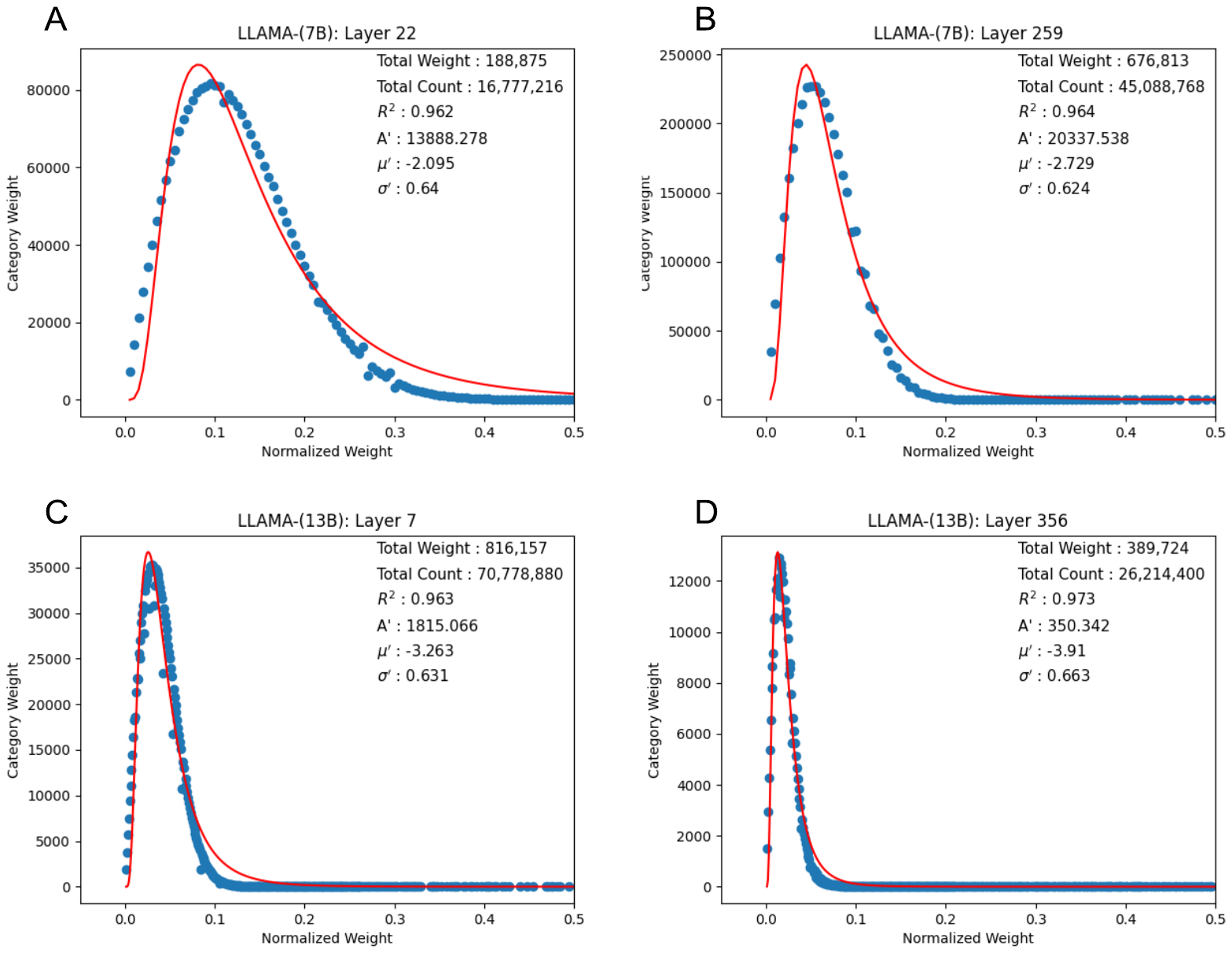}
    \caption{A \& B - Llama-2 7B; C \& D - Llama-2 13B}
    \label{fig:LLAMA-2}
\end{figure}

\newpage

%
%
%
%
%


\begin{table}[!ht]
    \centering
    \caption{Blazepose}
\begin{tabular}{|l|l|l|l|l|l|l|}
    
    \hline
        \#Layer & Total Count & Total Weight & A' & $\mu$' & $\sigma$' & $R^2$ \\ \hline
        27 & 36864 & 3493.07 & 3.753 & -1.695 & 0.788 & 0.903 \\ \hline
        29 & 36864 & 2800.54 & 2.950 & -2.046 & 0.750 & 0.94 \\ \hline
        31 & 36864 & 4162.39 & 4.436 & -1.674 & 0.742 & 0.929 \\ \hline
        33 & 36864 & 4661.61 & 5.004 & -1.572 & 0.724 & 0.921 \\ \hline
        35 & 36864 & 4620.21 & 4.926 & -1.587 & 0.727 & 0.92 \\ \hline
        37 & 55296 & 3789.97 & 3.970 & -2.118 & 0.733 & 0.955 \\ \hline
        39 & 82944 & 5640.63 & 5.936 & -2.106 & 0.764 & 0.953 \\
\hline
    \end{tabular}
    
\end{table}

    
    \begin{longtable}{|l|l|l|l|l|l|l|}
    \caption{BERT Small}
    \hline
        \#Layer & Total Count & Total Weight & A' & $\mu$' & $\sigma$' & $R^2$ \\ \hline
        1 & 2,34,40,896 & 9,56,970.79 & 1050.714 & -2.711 & 0.611 & 0.964 \\ \hline
        2 & 3,93,216 & 4,574.83 & 5.016 & -3.971 & 0.648 & 0.971 \\ \hline
        5 & 5,89,824 & 19,902.57 & 39.562 & -2.291 & 0.653 & 0.966 \\ \hline
        6 & 5,89,824 & 19,553.27 & 24.533 & -2.77 & 0.661 & 0.971 \\ \hline
        7 & 5,89,824 & 13,380.52 & 72.911 & -1.673 & 0.658 & 0.949 \\ \hline
        8 & 5,89,824 & 12,941.55 & 22.906 & -2.838 & 0.666 & 0.973 \\ \hline
        10 & 23,59,296 & 69,865.48 & 226.924 & -1.931 & 0.644 & 0.959 \\
\hline
    \end{longtable}
    

    \begin{longtable}{|l|l|l|l|l|l|l|}
        \caption{BERT Large}
    \hline
        \#Layer & Total Count & Total Weight & A' & $\mu$' & $\sigma$' & $R^2$ \\ \hline
        1 & 3,12,54,528 & 11,34,485.24 & 1050.674 & -2.99 & 0.638 & 0.967 \\ \hline
        2 & 5,24,288 & 6,600.64 & 6.577 & -3.993 & 0.638 & 0.969 \\ \hline
        5 & 10,48,576 & 29,041.93 & 60.741 & -2.44 & 0.645 & 0.964 \\ \hline
        6 & 10,48,576 & 28,589.24 & 41.765 & -2.816 & 0.651 & 0.968 \\ \hline
        7 & 10,48,576 & 21,569.71 & 145.17 & -1.562 & 0.657 & 0.953 \\ \hline
        8 & 10,48,576 & 20,306.56 & 21.478 & -3.488 & 0.666 & 0.975 \\ \hline
        10 & 41,94,304 & 1,15,362.42 & 295.501 & -2.242 & 0.647 & 0.961 \\
\hline
    \end{longtable}

    \begin{longtable}{|l|l|l|l|l|l|l|}
    \caption{LLAMA-2 7B}
    \hline
        \#Layer & Total Count & Total Weight & A' & $\mu$' & $\sigma$' & $R^2$ \\ \hline
        1 & 13,10,72,000 & 17,49,352.6 & 73424.31 & -2.498 & 0.629 & 0.959 \\ \hline
        2 & 1,67,77,216 & 1,21,611.4 & 1922.245 & -4.206 & 0.86 & 0.998 \\ \hline
        3 & 1,67,77,216 & 1,54,655.6 & 2330.648 & -3.959 & 0.845 & 0.998 \\ \hline
        4 & 1,67,77,216 & 1,34,305.5 & 10875.02 & -2.312 & 0.76 & 0.984 \\ \hline
        5 & 1,67,77,216 & 82,407.7 & 2257.873 & -4.038 & 0.726 & 0.997 \\ \hline
        6 & 4,50,88,768 & 5,80,491.9 & 8113.216 & -3.724 & 0.627 & 0.971 \\ \hline
        7 & 4,50,88,768 & 5,73,421.7 & 18806.59 & -2.809 & 0.627 & 0.967 \\
\hline
    \end{longtable}

    \begin{longtable}{|l|l|l|l|l|l|l|}
    \caption{LLAMA-2 13B}
    \hline
        \#Layer & Total Count & Total Weight & A' & $\mu$' & $\sigma$' & $R^2$ \\ \hline
        1 & 16,38,40,000 & 21,08,707.44 & 16206.96 & -1.928 & 0.7 & 0.826 \\ \hline
        2 & 2,62,14,400 & 1,43,007.45 & 189.704 & -4.369 & 1.05 & 0.996 \\ \hline
        3 & 2,62,14,400 & 1,74,317.13 & 243.006 & -4.1 & 1.045 & 0.995 \\ \hline
        4 & 2,62,14,400 & 1,87,058.83 & 876.438 & -2.957 & 0.77 & 0.969 \\ \hline
        5 & 2,62,14,400 & 1,32,453.39 & 286.956 & -4.114 & 0.744 & 0.99 \\ \hline
        6 & 7,07,78,880 & 8,34,811.67 & 1090.181 & -3.783 & 0.63 & 0.967 \\ \hline
        7 & 7,07,78,880 & 8,16,157.72 & 1815.066 & -3.263 & 0.631 & 0.963 \\
\hline
    \end{longtable}

\begin{table}[!ht]
    \caption{VGG16}
    \centering
    \begin{tabular}{|l|l|l|l|l|l|l|}
    \hline
        \#Layer & Total Count & Total Weight & A' & $\mu$' & $\sigma$' & $R^2$ \\ \hline
        1 & 1728 & 316.3954223 & 2.263 & 1.187 & 1.665 & 0.161 \\ \hline
        2 & 36864 & 1432.529306 & 2.772 & -2.114 & 0.841 & 0.962 \\ \hline
        3 & 73728 & 2634.214323 & 4.569 & -2.324 & 0.787 & 0.975 \\ \hline
        4 & 147456 & 4236.558286 & 12.153 & -2.061 & 0.736 & 0.974 \\ \hline
        5 & 294912 & 7038.238337 & 12.486 & -2.742 & 0.707 & 0.98 \\ \hline
        6 & 589824 & 11151.20801 & 30.102 & -2.562 & 0.675 & 0.974 \\ \hline
        7 & 589824 & 11405.32771 & 19.606 & -2.996 & 0.675 & 0.977 \\ \hline
        8 & 1179648 & 19328.59827 & 43.255 & -2.895 & 0.678 & 0.979 \\ \hline
        9 & 2359296 & 30011.00319 & 113.989 & -2.623 & 0.656 & 0.974 \\ \hline
        10 & 2359296 & 29925.61469 & 136.83 & -2.441 & 0.657 & 0.975 \\ \hline
        11 & 2359296 & 31940.58352 & 105.233 & -2.704 & 0.651 & 0.972 \\ \hline
        12 & 2359296 & 31653.64278 & 131.902 & -2.48 & 0.639 & 0.967 \\ \hline
        13 & 2359296 & 29488.95028 & 232.701 & -1.896 & 0.675 & 0.968 \\ \hline
        14 & 102760448 & 406312.2578 & 6910.126 & -2.283 & 0.666 & 0.968 \\ \hline
        15 & 16777216 & 133714.1187 & 2293.809 & -1.574 & 0.63 & 0.95 \\ \hline
        16 & 4096000 & 60237.04278 & 482.58 & -1.739 & 0.638 & 0.96 \\ \hline
    \end{tabular}
    
\end{table}

\begin{table}[!ht]
    \centering
    \caption{Xception}
    \begin{tabular}{|l|l|l|l|l|l|l|}
    \hline
        \#Layer & Total Count & Total Weight & A' & $\mu$' & $\sigma$' & $R^2$ \\ \hline
        27 & 1,86,368 & 10343.07 & 6.439 & -2.423 & 0.787 & 0.991 \\ \hline
        30 & 5,29,984 & 22656.47 & 9.914 & -2.696 & 0.741 & 0.988 \\ \hline
        32 & 1,86,368 & 7496.2 & 3.162 & -2.745 & 0.755 & 0.987 \\ \hline
        37 & 5,29,984 & 17238.02 & 7.464 & -2.943 & 0.773 & 0.989 \\ \hline
        40 & 5,29,984 & 18628.12 & 7.865 & -2.893 & 0.692 & 0.982 \\ \hline
        43 & 5,29,984 & 18816.25 & 8.042 & -2.884 & 0.715 & 0.985 \\ \hline
        47 & 5,29,984 & 14494.23 & 5.786 & -2.946 & 0.865 & 0.980 \\
\hline
    \end{tabular}
\end{table}